\documentstyle[12pt]{article}

\catcode`@=11
\def\secteqno{\@addtoreset{equation}{section}%
\def\theequation{\thesection.\arabic{equation}}}
\catcode`@=12
\secteqno
\newcommand{\be}{\begin{equation}}
\newcommand{\ee}{\end{equation}}
\newcommand{\bea}{\begin{eqnarray}}
\newcommand{\eea}{\end{eqnarray}}
\newcommand{\bref}[1]{(\ref{#1})}

\newcommand{\pa}{\partial}
\catcode`@=11

\begin{document}
\vfill
\vbox{
\today
\hfill 
}\null

\vskip 10mm
\begin{center}
{\Large\bf Comments on the Tetrad (Vielbeins).
\\
}\par
\vskip 10mm
{Takeshi FUKUYAMA\footnote{\tt E-mail:fukuyama@se.ritsumei.ac.jp}}
\par
\medskip
Department of Physics and R-GIRO, Ritsumeikan University,\\ 
Kusatsu, Shiga, 525-8577 Japan\\
\medskip
\vskip 10mm
\end{center}
\vskip 10mm
\begin{abstract}
We want to correct the misunderstandings on the tetrad (or veilbeins in general) appeared in many text books or review articles. The tetrad should be defined without any condition. $e_{\mu a}=\partial_\mu X_a$ with local Lorentz coordinates $X_a$ ia wrong in many sences: it gives the condition $\partial_\mu e_{\nu a}=\partial_\nu e_{\mu a}$, which leads us to the trivial result that the cyclic coefficients vanish identically and to the null Riemannian tensor.
Also $e_{\mu a}e_\nu^a=g_{\mu\nu}$ is not scalar under the local Lorentz transformation etc. We show how these deficits are remedied by the correct definition, $e_{\mu a}=D_\mu Z_a$ with local (Anti) de Sitter coordinates $Z_A$.

\end{abstract}

\vskip 2cm


From the gauge theoretical point of view \cite{Utiyama1}, one of the most important characters of gravity is the soldering of internal space of gauge symmetry of gravity with the external space. 
The other one is that gravity, at least the leading term at low energy, is linear in the Riemannian tensor unlike the other gauge theories.
So the gauge theory of gravity must reflects and explains these peculiarities.

In these processes, we can understand that the problem lies in the ambiguous situation of tetrad or metric in the gauge theoretical framework. There seems to exist some prejudice that the local symmetry of gravitation is Lorentz group (as the symmetry before the breaking) and misunderstanding that the tetrad is defined by
\be
e_{\mu}^a=\partial_\mu X^a(x)\equiv X^a,_\mu.
\label{derivative}
\ee
Here $X^a$ are the local Lorentz coordinates. 
The metric tensor is defined by
\be
g_{\mu\nu}=e_\mu^a e_\nu^b \eta_{ab},
\label{metric}
\ee
where $\eta_{ab}$ is the Minkowski metric.
Since there is no reason why the integrability condition does not hold,
\bref{derivative} forces $e_{\mu a}$ to be conditional,
\be
e_{\nu a,\mu}=e_{\mu a,\nu}.
\label{integral}
\ee
\bref{derivative} is often seen in many text books and popular review articles.
Unfortunately, it is absolutely wrong since tetrad defined by \bref{derivative} is not vector from the local Lorentz indicies and, therefore, $g_{\mu\nu}$ is not a scalar w.r.t. Lorents transformation:
\be
(X^U)^a=\left(U^{-1}(x)X(x)U(x)\right)^a
\ee
and
\be
(\partial_\mu X^U)^a\neq \left(U^{-1}\partial_\mu XU\right)^a
\ee
Therefore, the right-hand side of \bref{metric} is not a scalar under the local Lorentz transformation.

The pathology is easily seen by counting the physical degrees of freedom of $e_{\mu a}$: The tetrad should be subject only to the condition \bref{metric}
and the degrees of freedom of $e_{\mu a}$ are $16$. If \bref{integral} was correct, it forces us 10 constrains and $16-10=6$ degrees are left, leaving no room to invoke the Lorentz and diffeomorphism symmetries. The correct counting should be as follows. The invariance of Lorents transformation gives $6$ second class constraints and the diffeomorphism invariance does $4$ first class constraints \cite{Dirac}\cite{Fuku-Kami} and the physical degrees of freedom are $16-6-4\times 2=2$.

For pure gravity case, torsion free condition enables us to solve
$\omega_{\mu ab}$ in terms of the tetrad,
\be
\omega_{\mu ab}=-\frac{1}{2}e_\mu^c\left(\lambda_{abc}+\lambda_{bca}-\lambda_{cab}\right),
\label{omega}
\ee
where
\be
\lambda_{cab}=\left(e_{\rho c,\sigma}-e_{\sigma c,\rho}\right) e^\rho_ae^\sigma_b .
\label{lambda1}
\ee
If \bref{integral} was imposed on the tetrad, the above $\lambda_{cab}$ and therefore $\omega_{\mu ab}$ is identically vanish and spacetime must be flat.
This point will be shown more explicitly soon later.

$e_{\mu a}$ in the absence of Fermions is inversely described in terms of $g_{\mu\nu}$ as
\be
e_\mu^0=(-\sqrt{h},~\sqrt{h}g_i)~~e_\mu^{\bar{a}}=(0,e_i^{\bar{a}}).
\ee
Here 
\be
h=g_{00},~g_i=\frac{g_{0i}}{g_{00}},~e_i^{\bar{a}}e_{j\bar{a}}=g_{ij}-\frac{g_{0i}g_{0j}}{g_{00}}.
\ee
$\bar{a}$ and $(i,j)$ are the spatial parts of the Lorentz and world indicies, respectively.

The tetrad formulation is very powerful for the classification of homogeneou space \cite{L-L1}.

As we have said above, \bref{integral} gives rise to the problem that it leads us to the trivial results as follows \cite{L-L2}.
In the tetrad formalism of gravitation, the Ricci rotation coefficients play essential roles
\be
\gamma_{abc}\equiv e_{\mu a;\nu}e^\mu_be^\nu_c.
\ee
Here $e_{\mu a;\nu}$ is the covariant derivative w.r.t. the world coordinates,
\be
e_{\mu a;\nu}\equiv \pa_\nu e_{\mu a}-\Gamma_{\nu \mu}^\rho e_{\rho a}
\ee
and is distinguished from the covariant derivative w.r.t. the Lorentz coordinates $D_\nu e_{\mu a}$.
The relation between $\Gamma_{\nu \mu}^\rho$ and the spin connection will be given in 
\bref{metric cond}.

The linear combinations of the Ricci coefficients satisfies
\be
\gamma_{abc}-\gamma_{acb}=\lambda_{abc}
\label{lambda}
\ee
or inverse relations,
\be
\gamma_{abc}=\frac{1}{2}(\lambda_{abc}+\lambda_{bcd}-\lambda_{cab}),
\label{gamma}
\ee
are very important since the Riemannian tensor 
\bea
R_{abcd}&=&\left(e_{\mu a;\nu;\rho}-e_{\mu a;\rho;\nu}\right)e^\mu_be^\nu_ce^\rho_d\nonumber\\
&=& \gamma_{abc,d}-\gamma_{abd,c}+\gamma_{abf}(\gamma^f{}_{cd}-\gamma^f{}_{dc})+\gamma_{afc}\gamma^f{}_{bd}-\gamma_{afd}\gamma^f{}_{bc}
\eea
is described by $\lambda_{abc}$.
So if \bref{integral} was valid, $\lambda_{abc}$ and $\gamma_{abc}$ vanish identically and, therefore, the Riemannian tensors also vanish.
It should be remarked that we have assumed $\Gamma_{\mu\nu}^\rho$ to be symmetric on the lower indices.
This is not the necessary condition in every case but is valid in the case where we do not consider fermions in the first order formalism.
This point is discussed later in more detail.

Of course, if we consider the spinor, torsion may appear (in the first order formalism) and antisymmetric part of $\Gamma_{\mu\nu}^\rho$ may survive.
However, the tetrad formulation must be valid irrespectively to the presnce of spinor, and this pathology is very serious and \bref{integral}, and therefore \bref{derivative}, should be discarded.

Then, how to define the tetrad in place of \bref{derivative} ?
In order to answer to this question, it is necessary to consider the tetrad (vielbein in general) in the gauge framework \footnote{We can consider the metric as the gauge theory
of diffeomorphism \cite{F-U} though it can not incorporate Fermion.}.
Saying first the result, it is given by \bref{tetrad1} in the framework of
 SO(1,4) or SO(2,3) gauge theory \cite{Fukuyama} \cite{MacDowell} \cite{Stelle}. Let us explain it. In this theory, we consider, in place of the Lorentz group, the following local coordinates,
\bea
\sum_1^{d+1} Z_A^2&=&-l^2~~
\mbox{for SO(1,d) or}\\
\sum_1^{d+1} Z_A^2&=&l^2~~ 
\mbox{for SO(2,d-1)}.
\label{group}
\eea
Here we have described for general $d$ dimensional spacetime for later use.  Real $l$ measures the scale breaking from SO(2,d). Hereafter we discuss for SO(2,3), though this formalism is equally valid in case of SO(1,4).

Correspondng to SO(2,3), the covariant derivative is defined by
\be
D_\mu\psi=\left(\partial_\mu-i\omega_{\mu AB}S_{AB}/2\right)\psi~~(A,B=1,...,4,5).
\ee
Here $\omega_{\mu AB}$ are $4\times 10$ connection fields and $S_{AB}$ are the generators of anti de Sitter group.
You will see the differences of Poincare gauge theoris \cite{Kibble} in the subsequent arguments.
The field strength is derived from the commutation relation
\be
i[D_\mu,D_\nu]\psi=-\frac{1}{2}R_{\mu\nu AB}S_{AB}\psi.
\ee
\be
R_{\mu\nu AB}=\partial_\mu\omega_{\nu AB}-\partial_\nu\omega_{\mu AB}-\omega_{\mu AC}\omega_{\nu CB}+\omega_{\nu AC}\omega_{\mu CB}.
\ee
The Einstein's action is written as
\bea
I&=&\int d^4x \epsilon^{ABCDE}\epsilon^{\mu\nu\rho\sigma}(Z_A/l)\left[R_{\mu\nu BC}R_{\rho\sigma DE}/(16g^2)\right.\nonumber\\
&+& \left.D_\mu Z_BD_\nu Z_CD_\rho Z_DD_\sigma Z_E \sigma (x)\{(Z_A^2/l^2)-1\} \right].
\label{action1}
\eea

Here $\epsilon^{\mu\nu\lambda\sigma}$ and $\epsilon^{ABCDE}$ are fully antisymmetric tensors with $\epsilon^{0123}=1$ and $\epsilon^{12345}=1$, respectively.
It should be remarked that this action is a geometrical invariant and that we do not introduce metric ad hoc. 
After the gauge choice
\be
Z^A=(0,0,0,0,l),
\label{GF1}
\ee
\be
D_\mu Z_A=(\partial_\mu\delta_{AB}-\omega_{\mu AB})Z_B =\{\begin{array}{cc}-\omega_{\mu a5}l\equiv e_{\mu a} & \mbox{if}~ A=a\\
  0 & \mbox{if}~ A=5 \end{array} .
\label{tetrad1}
\ee
It is important that $e_{\mu a}$ transforms covariantly under the remaining 4-dim Lorentz rotation.
Generalized Riemannian tensor $R_{\mu\nu ab}$ is divided into two terms
\be
R_{\mu\nu ab}=\mathring{R}_{\mu\nu ab}-e_{[\mu a}e_{\nu]b}/l^2.
\ee
Here $\mathring{R}_{\mu\nu ab}$ is the conventional Riemannian tensor defined by
\be
\mathring{R}_{\mu\nu ab}=\partial_{[\mu}\omega_{\nu]ab}-\omega_{[\mu ac}\omega_{\nu]cb}
\ee
and $e_{[\mu a}e_{\nu]b}\equiv e_{\mu a}e_{\nu b}-e_{\nu a}e_{\mu b}$.

$L_{grav}$ takes the form of Euler class
\bea
L_{grav}&=&{\cal E}_4(gravity)= \epsilon^{abcd}\epsilon^{\mu\nu\rho\sigma}R_{\mu\nu ab}R_{\rho\sigma cd}/(16g^2)\nonumber\\
&=& \partial_\mu{\cal C}_4^\mu-e\left(\mathring{R}-\frac{6}{l^2}\right)/(16\pi G),
\label{4D2}
\eea
where 
\be 
16\pi G\equiv g^2l^2,
\ee
\be
e=\mbox{det}e_{\mu a},~~\mathring{R}_{\mu a}=e^{\nu b}\mathring{R}_{\mu\nu ab},~~\mathring{R}=e^{\mu a}\mathring{R}_{\mu a},
\ee
 and use has been made of
\bea
\epsilon^{abcd}\epsilon^{\mu\nu\rho\sigma}e_{\mu a}e_{\nu b}e_{\rho c}e_{\sigma d}&=&4! e,\nonumber\\
\epsilon^{abcd}\epsilon^{\mu\nu\rho\sigma}e_{\mu a}e_{\nu b}&=& 2e~e^{[\rho c}e^{\sigma] d}~~\mbox{etc.}
\eea
Here $e^{\mu a}e_{\mu b}=\delta_{ab},~e^{\mu a}e_{\nu a}=\delta^\mu_\nu$.
The quadratic term in $\mathring{R}_{\mu\nu ab}$ is total derivative $\partial_\mu {\cal C}_4^\mu$ (the Gauss-Bonnet term),
\be
\partial_\mu{\cal C}_4^\mu=e\left(\mathring{R}^2-4\mathring{R}_{\mu\nu}\mathring{R}^{\mu\nu}+\mathring{R}_{\mu\nu\rho\sigma}\mathring{R}^{\mu\nu\rho\sigma}\right)/(4g^2).
\label{CS}
\ee
This term with the definite coefficient $1/(4g^2)$ are both indispensable for the conservation of, for instance, mass and angular momentum of AdS Kerr Black hole \cite{Zanelli}.

So far we imposed \bref{GF1} before Euler variation of \bref{action1}.
This is not equal in general to imposing gauge condition after Euler variation which is the natural approach. This is easily understood if we replaced $\sigma$ term as $\sigma (x)\{(Z_A^2/l^2)-1\}^2$ from linear $\sigma (x)\{(Z_A^2/l^2)-1\}$ (See \bref{Eb}).
Let us explain in more detail \cite{Fuku-Kami}. We may express \bref{action1} in a coordinate-free form.
\be
I=\int\left[\frac{1}{2g^2}\tilde{\Theta}^{AB}\wedge\Theta_{AB}+\hat{\sigma}\left(\frac{Z_A^2}{l^2}-1\right)\right].
\ee
Here
\bea
\Theta_{AB}&\equiv& d\Omega_{AB}-\left(\Omega\wedge \Omega\right)_{AB}=\frac{1}{2}R_{\mu\nu AB}dx^\mu\wedge dx^\nu,\nonumber\\
\Omega_{AB}&\equiv& \omega_{\mu AB}dx^\mu,\nonumber\\
\tilde{\Theta}^{AB}&\equiv& \frac{1}{2}\epsilon^{ABCDE}\frac{Z_C}{l}\Theta_{DE},\\
\hat{\sigma}&=&\epsilon^{ABCDE}\epsilon^{\mu\nu\rho\sigma}(Z_A/l)D_\mu Z_BD_\nu Z_CD_\rho Z_DD_\sigma Z_E\sigma d^4x.\nonumber
\eea
The Euler equations are derived by taking the variation of the action with respect to $\Omega,Z$ and $\sigma$:
\bea
d\tilde{\Theta}^{AB}-[\Omega,\tilde{\Theta}]^{AB}=0,\label{Ea}\\
\epsilon^{ABCDE}\frac{1}{4g^2l}\Theta_{BC}\Theta_{DE}+\frac{2}{l^2}Z^A\hat{\sigma}=0,\label{Eb}\\
\frac{Z_A^2}{l^2}-1=0.\label{Ec}
\eea
\bref{Eb} can be regarded as the equation determining $\sigma$ in terms of connections.
In the gauge \bref{GF1}, \bref{Ea} becomes
\bea
d\tilde{\Theta}^{ab}-[\Omega,\tilde{\Theta}]^{ab}=0~~~~\mbox{for}~(AB)=ab,
\label{Ed}\\
\tilde{\Theta}^{ab}\Omega^{b5}=0~~~~\mbox{for}~(AB)=a5.
\label{Ee}
\eea
In \bref{Ed} summation is taken among small Latin.
It should be remarked that \bref{Ed} and \bref{Ee} are dual to the Bianchi and cyclic identities, respectively. \bref{Ed} correspond to the first order formalism of Palatini \cite{Palatini}.
\bref{Ee} is the Einstein equation in vacuum. 
Thus the arguments \bref{4D2}-\bref{CS} are remained valid.

If we perform exterior derivative $d$ on \bref{Ee} and use \bref{Ee} again, we obtain
\be
\tilde{\Theta}^{ab}(d\Omega^{b5}-\Omega^{bc}\Omega^{c5})=0.
\ee
As is easily seen antisymmetric $\tilde{\Theta}^{ab}$ is nondegenerate in \bref{Ee} and we obtain torsion-free condition $\Theta^{a5}=0$ or explicitly
\be
R_{\mu\nu a5}=-\frac{1}{l}\left(\pa_\mu e_{\nu a}-\pa_\nu e_{\mu a}-\omega_{\mu ac}e_\nu^c+\omega_{\nu ac}e_\mu^c\right)=0.
\label{torsion}
\ee

Furthermore, we use the metric condition of $e_{\mu a}$
\be
e^a_{\mu|\nu}\equiv \pa_\nu e_{\mu a}-\omega_{\mu ac}e_\nu^c-\Gamma_{\nu\mu}^\rho e_{\rho a}=0.
\label{metric cond}
\ee
\bref{torsion} represents
\be
\Gamma_{\mu\nu}^\rho=\Gamma_{\nu\mu}^\rho.
\ee
Fermions (Dirac, Weyl, Majorana) are studied in the framework of (Anti) de Sitter gravity in \cite{Ikeda}.

\bref{4D2} shows the reason why the gravitational action is linear on the Riemannian tensor, and \bref{tetrad1} indicates that it transforms as vector under the local Lorentz transfomation. The tetrad is not subject to any condition since it comes from the additional gauge freedom of $\omega_{5a}$. Hamilton formulation of full (Anti) deSitter gravity in terms of Dirac prescription \cite{Dirac} was performed in \cite{Fuku-Kami} and physical degrees of freedom are correctly 2.
That is, the total degrees of freedom of the system are $92$ ($40~\omega_{\mu AB},5~Z_A,\sigma$ and their conjugate momenta). The number of first class constraints is $20$,
\bea
\pi^{0BC}&=&0,\\
D_i\pi^{iBC}&=&0,
\eea
and that of second class constraints is $48$
\bea
&&\pi^{ibc}-\frac{1}{2g^2}\epsilon^{ijk}\epsilon^{bcde}R_{jkde}=0,\\
&&\pi^{ia5}=0,\\
&&p^A=0,\\
&&Z_a=0,~Z_5=l,\\
&&p_\sigma=0,\\
&&\sigma-\frac{1}{8g^2}\epsilon^{bcde}\epsilon^{ijk}R_{jkde}\lambda^0_{ibc}=0,\\
&&R_{jka5}=0,
\label{last}
\eea
where $\pi^{\mu AB},p^A,p_\sigma$ are canonical momenta of $\omega_{\mu AB},Z_A,\hat{\sigma}$, respectively. $\lambda^0_{ibc}$ are the Lagrange multiplier of $\pi^{ibc}$. Only $6$ equations of \bref{last} are independent (See the detail \cite{Fuku-Kami}). So we have $4 (=92-2\times 20-48$) physical degrees of freedom in Hamilton formalism in agreement with the gravitational field.

Thus the problems discussed above have been all solved.

Finally we comment on the vielbein in the other dimensions than four.
Our frmulation is straihtforwardly applied to 3,5,6,.. dimensions \cite{Fukuyama3} \footnote{For two dimensional case we need the special treatment concerning with conformal invariance \cite{Fukuyama2}}.
It is enough to discuss five dimensions for the present paper's purpose.
Usually, Euler class can be defined only in even dimensions, However, in our formulation there is no essential difference between odd and even dimensions.
Indeed, this formulation is easily extended to five dimensional spacetime.
That is
\bea
I&=&\int d^5x \epsilon^{ABCDEF}\epsilon^{\mu\nu\rho\sigma\lambda}(Z_A/l)D_\mu Z_B\left[R_{\nu\rho CD}R_{\sigma\lambda EF}/(48g^2l)\right.\nonumber\\
&+& \left. D_\nu Z_CD_\rho Z_DD_\sigma Z_ED_\lambda Z_F\sigma (x)\sum_{A=1}^6\{(Z_A^2/l^2)-1\} \right]
\label{5D}
\eea
with 
\be
Z_A=(0,0,0,0,0,l).
\label{GF2}
\ee
In this case
\be
D_\mu Z_A=(\partial_\mu\delta_{AB}-\omega_{\mu AB})Z_B =\{\begin{array}{cc}-\omega_{\mu a6}l=e_{\mu a} & \mbox{if}~ A=a\\
  0 & \mbox{if}~ A=6 \end{array} ,
\ee
Here $\mu$ and $a$ run over 1,..,5 in world and local Lorentz coordinates, respectively.
Consequenly \bref{5D} is reduced to
\bea
L_{grav}&=& \epsilon^{abcde}\epsilon^{\mu\nu\rho\sigma\lambda}e_{\mu a}R_{\nu\rho bc}R_{\sigma\lambda de}/(48g^2l)\nonumber\\
&=&\epsilon^{abcde}\epsilon^{\mu\nu\rho\sigma\lambda}e_{\mu a}\mathring{R}_{\nu\rho bc}\mathring{R}_{\sigma\lambda de}/(48g^2l) -e\left(\mathring{R}-\frac{10}{l^2}\right)/(16\pi G_5),
\label{d3}
\eea
where $G_5$ is five dimensional Newton constant.
Thus we obtain $AdS_5$ in low enery scale. In this case, however, higer derivative terms (the first term of \bref{d3}) are not total derivatives and change the equation of motion in high enery region and do therefore Black Hole solution and its near horizon property.
The explicit representation of the first term of \bref{d3} is
\be
\epsilon^{abcde}\epsilon^{\mu\nu\rho\sigma\lambda}e_{\mu a}R_{\nu\rho bc}R_{\sigma\lambda de}/(48g^2l)=e\left(\mathring{R}^2-4\mathring{R}_{\mu\nu}\mathring{R}^{\mu\nu}+\mathring{R}_{\mu\nu\rho\sigma}\mathring{R}^{\mu\nu\rho\sigma}\right)/(12g^2l).
\ee
The coefficient $12g^2l$ is fixed by five dimensional gravitational constant $G_5$ and cosmological constant $\Lambda_5$,
\be
16\pi G_5=g^2l^3,~~\Lambda_5=-5/l^2.
\ee

For six and higher dimensions there appears cubic term on the Riemannian tensor. Even in that case all coefficients are definitely given in terms of $G$ and $l$.
This is quite different from the conventional approaches of Lovelock Lagrangian
\cite{Lovelock}. The coefficients of quadratic and cubic terms are each left free parameters there. However, as I have emphasized, the special coeffient determined by our theory has the special merit for four dimensions and we may expect the same thing occurs in the other dimensions, which is now under investigation.

\section*{Acknowledgments}
We would like to thank A.Randono for very useful conversations. 
This work is supported in part by the grant-in-Aid 
for Scientific Research from the Ministry of Education, 
Science and Culture of Japan (No. 20540282).

\end{document}